\documentclass[letter,prd,reprint,superscriptaddress,
unsortedaddress]{revtex4-1}
\usepackage[utf8x]{inputenc}
\usepackage{amssymb,graphicx}
\usepackage[english]{babel}
\usepackage{amsmath}
\usepackage{amsfonts}
\usepackage{amssymb}
\usepackage{fullpage}
\usepackage{ifpdf}

\def\e{{\rm e}}

\newcommand{\be}{\begin{equation}}
\newcommand{\ee}{\end{equation}}
\newcommand{\bea}{\begin{eqnarray}}
\newcommand{\eea}{\end{eqnarray}}
\newcommand{\bg}{\begin{gather}}
\newcommand{\eg}{\end{gather}}
\newcommand{\bseq}{\begin{subequations}}
\newcommand{\eseq}{\end{subequations}}

\newcommand{\ch}{\mathop{\rm ch}\nolimits}
\renewcommand{\ln}{\mathop{\rm ln}\nolimits}

\def\e{{\rm e}}

\def\ch{{\rm ch}}
\def\sh{{\rm sh}}

\def\g{{\mathrm{g}}}
\def\ln{{\mathrm{ln}}}

\begin{document}
\title{IR properties of one loop corrections to 
  brane-to-brane propagators in
  models with localized vector bosons}
\author{D.V.~Kirpichnikov}
\affiliation{Institute for Nuclear  Research of the Russian Academy 
of Sciences, Prospect of the 60th Anniversary of October 7a, 
Moscow, Russia, 117312}

\begin{abstract}
We discuss the one loop effects
 of massless fermion fields on the low energy
 vector brane-to-brane propagators in the framework 
 of two QED brane-world scenarios.  
 We  show that  one loop photon brane-to-brane 
 propagator has a power law pathologic
 IR  divergences  
 in the 5D QED brane-world model with mass gap
between the vector zero mode and continuous states.
We also find that  bulk fermions 
 do not give rise to IR divergences in a photon
 brane-to-brane Green's function at least at the 
 one loop level 
 in the framework of 6D QED brane model 
 with gapless mass spectrum between vector zero mode
 and higher states.
\end{abstract}

\maketitle
\section{Introduction}

Localizing scalars and fermions in brane-world
field theoretic models is conceptually straightforward
\cite{Rubakov:1983bb, 
Akama:1982jy,RandjbarDaemi:2000cr}. On the other hand,
gauge field localization is tricky. One either does 
not allow gauge fields to penetrate into the bulk 
\cite{Dvali:1996xe}, or has to deal with charge universality
\cite{Dubovsky:2001pe}, i.~e., the property that the 
effective 4D gauge coupling must be independent
of the bulk wave function of a charged
particle. The latter property implies that the gauge 
field zero mode is constant along extra dimensions 
(unlike zero modes of scalars and fermions, which typically 
decay away from the brane).

\begin{figure}
\begin{center}
\includegraphics[width=0.5\textwidth]{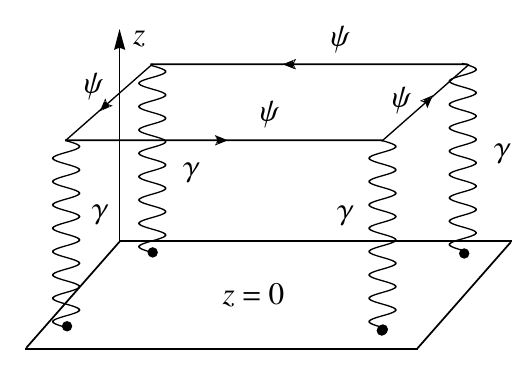}
\caption {Light by light scattering amplitude 
in a domain wall set up,
$\mathcal{M}(\gamma\gamma\!
\rightarrow\!\gamma\gamma)\!
\propto\!L \rightarrow \infty$.
 \label{figure1}}
\end{center}
\end{figure}

The independence of the gauge field zero mode of 
extra-dimensional coordinates may lead to infrared 
problems. Indeed, it has been pointed out in
Refs.~\cite{Smolyakov:2011hv, Smolyakov:2012ud}  
that charged bulk fields induce one-loop
scattering amplitude of the zero mode gauge bosons, 
see Fig.~\ref{figure1},
which diverges in 5D lineary with the size of extra
dimension $L$, provided that 
the effective mass of the charged field does not 
grow indefinitely away from the brane. This is easy 
to understand: suppose one 
keeps the size of the loop finite and moves it towards
$z \rightarrow \infty$, where $z$ is the extra coordinate.
Since the gauge field zero mode is constant in $z$, 
and the effective mass of the charged field is constant
at large $z$, the loop contribution is independent of
$z$, and the integration over $z$ gives the volume
of extra dimension. Likewise, the one-loop correction
to the 4D propagator of the gauge field 
boson in the zero mode state
 is also proportional to~$L$. 
 
 This observation,
however, does not necessarily mean that a theory has 
unacceptable IR pathologies. Indeed, since the zero mode
is strongly non-local along extra dimension, it 
cannot be produced alone by a local source. 
The IR pathology is indeed likely to be there in 
models with a gap between the zero and higher modes in the 
gauge field sector \cite{Smolyakov:2011hv}: at low 
4D momenta heavy states decouple, and the only relevant 
degree of freedom is the zero mode. However,
the gap is absent in other models of gauge field 
localization, notably, generalizations  
\cite{Oda:2000zc, Gherghetta:2000jf,Dubovsky:2000av}
 of the RSII set up
 \cite{Randall:1999vf}. Arguments have been given 
 some time ago \cite{Dubovsky:2002xv}, suggesting that 
 models of the latter type may be free of IR
 pathologies.

 In this paper we calculate and compare
 objects of direct physical significance in models with 
 and without gap. Since in the brane-world scenario
 one is interested primarily in the processes with particles
 (say, charged fermions) residing on the brane, the 
 object of  particular relevance is 
 the brane-to-brane gauge field propagator. As
 we are interested in IR properties, we are 
 going to study its behavior as $p\rightarrow 0$,
 where $p$ is 4D momentum. We will calculate corrections 
 to  the brane-to-brane gauge field 
 propagators, which are due to fermion loops in the bulk; 
 we consider massless fermions
 for simplicity.  For concreteness, we consider models
 of Ref.~\cite{Smolyakov:2011hv} (with the gap) and 
 Ref.~\cite{Dubovsky:2000av} with $n=1$ (gapless).
Our results confirm that the models with the gap
 between the
gauge zero mode and higher modes are IR pathological.
Namely, we will see that one loop correction to the 
brane-to-brane gauge field propagator behaves as 
$1/p^3$ at low 4D momenta (the bare propagator
is proportional to $1/p^2$, as usual). On the other 
hand, the one loop correction in the gapless model is free
of pathology: the one-loop correction behaves as $1/p^2$,
so it merely introduces the wave function renormalization.

The organization of this paper is as follows. 
In Sec.~\ref{DomainWallAction} 
we consider 5D spinor QED with gauge 
field localized on the domain wall. We 
introduce the model in Sec.~\ref{theDWmod}
and 
in Sec.~\ref{DomainWallCorrections} we explicitly 
calculate one loop fermion 
correction to the vector brane-to-brane propagator.
 In Sec.~\ref{RSII-1}  we consider 6D spinor 
QED in the background of RSII-1 metric with
one compact extra dimension $\theta$ and one 
infinite extra dimension $z$.
We calculate the vector propagator in 
Sec.~\ref{VecGreenRSII}. 
In Sec.~\ref{mslsRSII} we derive one loop 
contribution of $\theta$-homogeneous KK fermions
 into the vector brane-to-brane propagator. 
In Sec.~\ref{KKRSII} we discuss a
position depenent cutoff scheme
 for calculating the 
one loop contribution of $\theta$-inhomogeneous 
 KK excitations of fermions to the vector propagator,
 and calculate this correction explicitly.
 We conclude in Sec.~\ref{discussion}. Technical 
 details are collected in Appendices.

\section{Domain wall set up
\label{DomainWallAction}}
\subsection{The model \label{theDWmod}}
In this section we consider Euclidean 5D brane-world model
 with fermion and vector fields propagating in the bulk
\cite{Smolyakov:2011hv}. 
 The action of the 
model is  
 \begin{equation}
S\!=\!\!
\int\!\! d^4x\, dz \Bigl[ 
\,\frac{1}{4}\phi^2(z)F_{MN}^2+ i\overline{\Psi}
\Gamma^M(\partial_M-i\g_5 A_M)\Psi \Bigr],
\label{DWaction}
\end{equation}
where indices $M,N$ label 5D space, $M,N=0,1,2,3,5$,
and
\begin{equation}
\phi(z)=1/\ch \alpha z
\end{equation}
 is a field configuration which ensures the localization of
 vector zero mode on the brane;
 the parameter  $1/\alpha$ is related 
 to the brane thickness. This mechanism of gauge field localization 
 on a higher dimensional deffect
 is analogous to those considered in Ref. \cite{Barnaveli:1990bi}.
  For simplicity
we assume  that the fermions propagate in the flat 5D
bulk. 5D and 4D couplings are related by
\begin{equation}
\g_5=\frac{1}{\sqrt{\alpha}}\g_4.
\label{g5g4}
\end{equation} 
It is convenient to introduce the new field $B_M$,
 which is related to $A_M$ as follows:
\begin{equation}
B_M= \phi A_M.
\label{VecRescKink}
\end{equation}
The Lagrangian of the vector field $B_M$ can be
 written as follows:
\begin{equation}
\begin{split}
\mathcal{L}\!=\!\frac{1}{2} B_{\mu}\!\left[
 \eta_{\mu\nu}\!\left(\!-\partial^2_\lambda\!-\!
 \partial^2_z\!+\!\frac{\phi''}{\phi}
 \right)\!\!+\!
 \partial_\mu\partial_\nu \right]\!\!B_\nu-
 \\
 -\frac{1}{2}B_z\partial^2_\mu B_z-
 B_z\!\left(\frac{\phi'}{\phi}-
 \partial_z\right)\!\!\partial_\mu B_\mu,
\label{VecActResc}
\end{split}
 \end{equation}
 where Greek indices refer to 4D,~$\mu,\nu=0,1,2,3$.
We set the gauge  $B_z=0$, then the vector field 
$B_\mu$ is transverse,
$\partial_\mu B_\mu=0$.
 KK modes $B^{(m)}(z)$ of the vector field $B_\mu$  
 obey
 \begin{figure}[t]
\begin{center}
\includegraphics[width=0.5\textwidth]{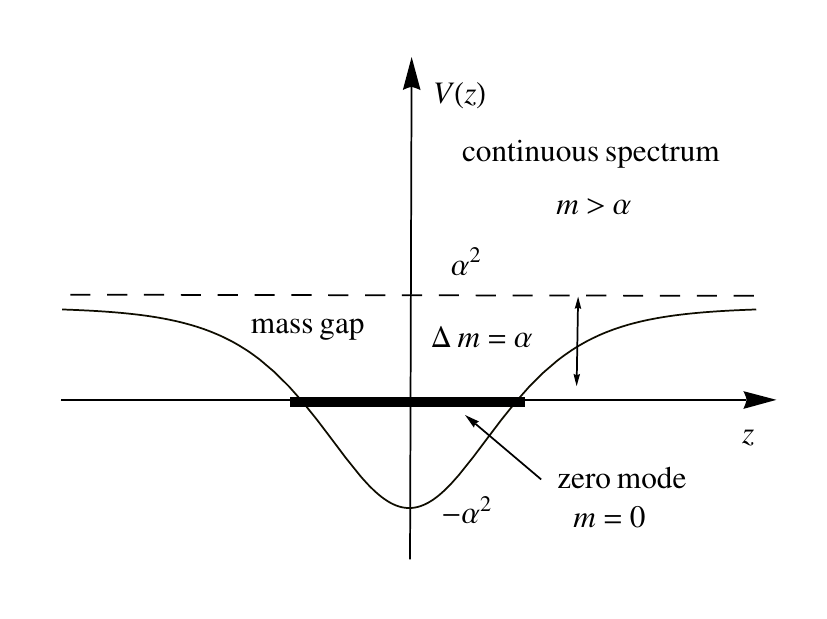}
\caption {Domain wall effective potential.
 \label{figure3}}
\end{center}
\end{figure} 
\begin{equation}
\left(-\partial^2_z+\frac{\phi''}{\phi}
\right)B^{(m)}(z)= m^2 B^{(m)}(z).
\label{VecEq1}
\end{equation}
where $m$ is 4D mass. 
The mass spectrum is determined by the
 quantum-mechanical potential 
 $$
 V(z)=\frac{\phi''}{\phi} = \alpha^2-
 \frac{2 \alpha^2}{\ch^2 \alpha z}.
 $$ 
In this potential the vector field 
has one bound state~(see Fig.~\ref{figure3}), 
which is actually the zero mode,
\begin{equation}
 B^{(0)}(z)= \sqrt{\frac{\alpha}{2}}
 \frac{1}{\ch\, \alpha z}.
 \label{DWZeroMode}
 \end{equation}
This mode is normalized with unit measure,
in accordance with the
Lagrangian~(\ref{VecActResc}). Eq.~(\ref{VecEq1}) 
has also solutions $B^{(m)}(z)$, which correspond to 
the continuous spectrum starting from  $m=\alpha$.
 Therefore,
the zero mode $B^{(0)}(z)$ is separated from higher
modes $B^{(m)}(z)$ by non-zero mass  gap $\Delta m = \alpha$. 

Now let us consider the propagator of the vector 
field $B_\mu$ from brane to bulk,
$G^B_{\mu\nu}(p,z,0)=\langle B_{\mu}(p,z) 
B_{\nu}(p,0)\rangle $.
This propagator  obeys
\begin{equation}
\left(p^2-\partial^2_z+\alpha^2-\frac{2 \alpha^2}{\ch^2 \alpha z}\right) G^B_{\mu\nu}(p,z,0)=\eta_{\mu\nu}\delta(z). 
\label{kinkVecGrF2}
\end{equation} 
The solution to 
Eq.~(\ref{kinkVecGrF2}) is
$G^B_{\mu\nu}(p,z,0)=\eta_{\mu\nu}G_B(p,z,0)$,
where (see Appendix~\ref{GreeDW})
\begin{equation}
 G_B(p,z,0)=
 \frac{1}{4\,\ch \alpha z}\left(\frac{\e^{-(\chi-\alpha)|z|}}
 {\chi-\alpha}+
 \frac{\e^{-(\chi+\alpha)|z|}}{\chi+\alpha}\right),
\label{GRenormKink}
\end{equation}
and 
\begin{equation}
\chi=\sqrt{p^2+\alpha^2}.
\label{chi}
\end{equation}
It is worth noting that brane-to-bulk propagator
of the vector field $A_{\mu}(p,z)$ is
\begin{equation}
\begin{split}
\langle A_{\mu}(p,z)A_{\nu}(p,0) \rangle
&=\eta_{\mu\nu}G_B(p,z,0)/\phi(z)
\\
&\equiv
\eta_{\mu\nu} G_A(p,z,0),
\label{VecGFDWAdef}
\end{split}
\end{equation}
where 
\begin{equation}
 G_A(p,z,0)=
 \frac{1}{4}\left(\frac{\e^{-(\chi-\alpha)|z|}}
 {\chi-\alpha}+
 \frac{\e^{-(\chi+\alpha)|z|}}{\chi+\alpha}\right).
\label{GRenormKinkAterm}
\end{equation}

Vector fields $A_M$ and $B_M$ coincide on the brane $z=0$,
their propagators from brane to brane coincide as well
\begin{equation}
G_A(p,0,0)=G_B(p,0,0)=\chi/(2p^2).
\label{zeromodeVecKink}
\end{equation}
At low energy, $p\ll \alpha$, only zero mode of gauge field
is relevant, 
and the propagator (\ref{zeromodeVecKink}) takes the form
\begin{equation}
G_A(p,0,0)=\alpha/(2 p^2)
\label{zeromodeVecKink2}.
\end{equation}
It follows from Eqs.~(\ref{chi}) 
and~(\ref{GRenormKinkAterm}) that at small $p$
\begin{equation}
G_{A}(p,z,0)= \frac{\alpha}{2p^2}
\exp{\left(-\frac{p^2 |z|}{2\alpha}\right)},
\label{VecGFDWA}
\end{equation}
 hence the bulk vector Green's function 
$G_{A}(p,z,0)$  decreases slowly towards 
$z\rightarrow \infty$ in the IR regime.
We will
 also
use the propagator $G_A(p,z,0)$ in the momentum space
\begin{equation}
\begin{split}
&\widetilde{G}_A(p,p_z)=\int\limits_{-\infty}^{+\infty}
\!
dz\,  G_A(p,z,0)\, \e^{i p_z z}
\\
&= \frac{1}{2}\left(\frac{1}{(\chi\!-\!\alpha)^2\!+\!p_z^2}+
\frac{1}{(\chi\!+\!\alpha)^2\!+\!p_z^2}\right).
\end{split}
\label{GreenKinkFourier}
\end{equation}
We find from Eqs.~(\ref{chi})
 and~(\ref{GreenKinkFourier}),
that as~$p_z\rightarrow~0$ and
 $p/\alpha \ll 1$, then the brane-to-bulk
 vector propagator in the momentum space
  tends to    
\begin{equation}
\widetilde{G}_A(p,p_z)= \frac{1}{2}\frac{1}{(p_z^2+p^4/(4\alpha^2))}
\label{FourierDW}. 
\end{equation}
We note that Eq.~(\ref{FourierDW}) differs considerably 
from the vector propagator in the flat 5D space
\begin{equation}
\widetilde{G}^{flat}_A(p,p_z)=\frac{1}{p^2+p_z^2}.
\end{equation}
In Sec.~\ref{DomainWallCorrections} we show 
that the term $p^4/(4\alpha^2)$ in Eq.~(\ref{FourierDW})
 leads to IR pathology of the one-loop vector 
 brane-to-brane propagator.

\subsection{One loop fermion contribution 
to the vector brane-to-brane propagator
 \label{DomainWallCorrections}}

 \begin{figure}[h]
\begin{center}
\includegraphics[width=0.47\textwidth]{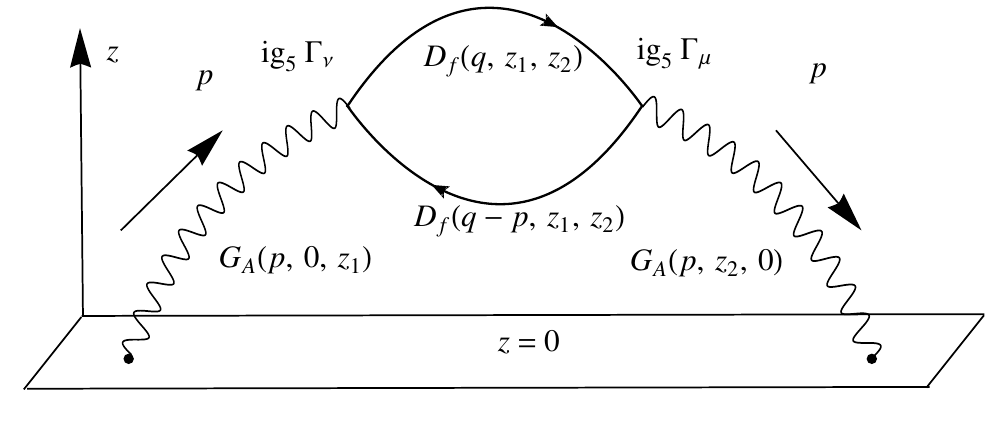}
\caption {One loop correction to the 
vector brane-to-brane propagator.
 \label{figure2}}
\end{center}
\end{figure} 
In this section we calculate the one loop
fermion contribution to the vector propagator 
from brane to brane (see Fig.~\ref{figure2}). The  coupling 
of the fields $A_\mu$ and  $\Psi$ is
\begin{equation}
S_{int}[\Psi, A]= 
 \int d^4 x\, dz\,  \g_5   \overline{\Psi} \,
\Gamma^\mu A_{\mu}\Psi.
\label{IntDomainWall}
\end{equation} 
One loop brane-to-brane vector Green's function is
\begin{equation}
\mathcal{G}_{\mu\nu}(p)= 
G^{(0)}_{\mu\nu}(p)+G_{\mu\nu}^{(1)}(p),
\end{equation} 
where tree level propagator is $G^{(0)}_{\mu\nu}(p)\equiv 
\eta_{\mu\nu} G_A(p,0,0)$, and one loop contribution
$G_{\mu\nu}^{(1)}(p)$ 
is given by 
\begin{equation}
\begin{split}
&G_{\mu\nu}^{(1)}(p)\!=\!(i \g_5)^2 
\!\!\!\int\limits_{-\infty}^{\infty}\!\!  
dz_1\!
\!\!\int\limits_{-\infty}^{\infty} \!\!\!dz_2\,
 G_A(p,z_1,0)G_A(p,z_2,0)\times
\\
&\times\!\!\int\!\!
\frac{d^4 q}{(2\pi)^4}\!(-1)
\mathrm{Tr} \left[\Gamma_\mu D_f(q-p,z_1,z_2)
\Gamma_\nu D_f(q,z_1,z_2)  \right].
\label{LoopCorrRSII-1}
\end{split}
\end{equation} 
Green's function of massless 5D  
fermions  $D_f(q,z_1,z_2)$ 
has the form
\begin{equation}
D_f(q,z,z')=\int \limits_{-\infty}^{\infty}
 \frac{dq_z}{2\pi} 
\frac{\widehat{Q}}{Q^2}
\,\e^{i q_z (z-z')},
\label{FermMassiveGreenFourier}
\end{equation}
where 
$\widehat{Q}=\Gamma^M q_M=\Gamma^{\mu}q_\mu+\Gamma^z q_z$, 
$Q^2$ is 5D momentum squared, $Q^2=q^2+q_z^2$. 
Therefore,  the one loop fermion 
contribution (\ref{LoopCorrRSII-1})
 can be written as follows
\begin{equation}
G^{(1)}_{\mu\nu}(p)\!=\!(\g_5)^2 \!\!\!  
\int\limits_{-\infty}^\infty \frac{dp_z}{2\pi}
\left|\widetilde{G}_A(p,p_z)\right|^2 
\widetilde{\Pi}_{\mu\nu}(p,p_z)
\label{LoopCorrection11}.
\end{equation}
where $\widetilde{G}_A(p,p_z)$ is 
given by Eq.~(\ref{GreenKinkFourier}) and 
$\widetilde{\Pi}_{\mu\nu}$ is vacuum 
polarization operator 
in flat 5D space with 4D indices:
\begin{equation}
 \widetilde{\Pi}_{\mu\nu}(p,p_z)=
\int\!\! \frac{d^5Q}{(2\pi)^5}
\frac{{\tt Tr}\Bigl[\Gamma_\mu \widehat{Q}
\Gamma_\nu  (\widehat{Q}-\widehat{P}) \Bigr]}
{Q^2(Q-P)^2}.\label{PiabtildeKink}
\end{equation}
 We make use of dimensional regularization and
 obtain 
\begin{equation}
\widetilde{\Pi}_{\mu\nu}(p,p_z)=\frac{3\pi}{64}\,
\frac{2^2}{(4\pi)^2}P
\Bigl(\eta_{\mu\nu}P^2-p_\mu p_\nu\Bigr).
\label{PiABFourier}
\end{equation}
where 5D momentum squared is
\begin{equation}
P^2=p^2+p_z^2.
\label{5dPsquared}
\end{equation}
Let us consider the one loop correction 
$G^{(1)}_{\mu\nu}(p)$
 in the low energy limit, $p\rightarrow 0$. 
We substitute 
Eqs.~(\ref{GreenKinkFourier}) and~(\ref{PiABFourier}) into 
 Eq.~(\ref{LoopCorrection11}) and integrate 
 (\ref{LoopCorrection11}) over $p_z$, then we get
\begin{equation}
\begin{split}
G^{(1)}_{\mu\nu}(p)\!=& \frac{(g_4)^2}{2\pi}
\frac{3\pi}{64}\frac{2^2}{(4\pi)^2} \frac{\alpha}{2p^2}
\Big[
\left(\!\eta_{\mu\nu}\!-\!\frac{p_\mu p_\nu}{p^2}\right)\!
Q_T(p)+
\\
&+\frac{p_\mu p_\nu}{p^2}Q_{L}(p)
\Big].
\label{PiABLoopCorrKink}
\end{split}
\end{equation}
The factor $\alpha/(2p^2)$ in
 Eq.~(\ref{PiABLoopCorrKink})  
is due to the vector zero mode 
 (see Eq.~(\ref{zeromodeVecKink2})). 
Functions $Q_T(p)$
 and $Q_L(p)$ are defined by 
\begin{equation}
\begin{split}
Q_T(p)\!=\!\!\int\limits_{-\infty}^{+\infty}\!\!
dp_z \frac{2p^2}{\alpha^2} P^3 
\Bigl|\widetilde{G}_A(p,p_z)\Bigr|^2, 
\\
Q_L(p)\!=\!\!\int\limits_{-\infty}^{+\infty}
\!\!
dp_z \frac{2p^2}{\alpha^2} P p_z^2 
\Bigl|\widetilde{G}_A(p,p_z)\Bigr|^2.
\label{QsDW}
\end{split}
\end{equation} 
One obtains in IR regime $p \rightarrow 0$ 
\begin{equation}
\begin{split}
&Q_T(p)=\frac{2\pi \alpha}{p}+
\frac{9\pi p}{4\alpha}+
\\
&+\frac{p^2}{\alpha^2}\left(4\, \ln\frac{ 2\Lambda}{p}
+3\ln\frac{p}{4\alpha}
 -\frac{11}{6}\right)+
 O\left(\frac{p^3}{\alpha^3}\right)
\label{PiTransverse}
\end{split}
\end{equation}
\begin{equation}
Q_L(p)\!=\!\frac{\pi p}{2\alpha}\!+\!
\frac{p^2}{\alpha^2}\!\left(\!4 \ln\frac{ 2\Lambda}{p}
\!+\!3\ln\frac{p}{4\alpha}
 \!-\!\frac{3}{2}\right)\!+\!
 O\!\left(\!\frac{p^3}{\alpha^3}\!\right),
\label{PiLongitude}
\end{equation}
where $\Lambda$ is the UV cutoff scale.
It follows from Eq.~(\ref{PiTransverse}) that
in the low energy limit
  $p/\alpha \ll 1$
the function $Q_T(p)$ is proportional to $\alpha/p $. 
   The reason for this singularity is as follows.
 We have pointed out in Sec.~\ref{theDWmod} 
 that brane-to-bulk propagator $\widetilde{G}_A(p,p_z)$ 
 has a peculiar term $p^4/(4\alpha^2)$ in the denominator
 as $p_z \rightarrow 0$ 
 (see Eq.~(\ref{FourierDW})). This means that the
 integral~(\ref{QsDW}) for $Q_{T}(q)$ has the following
 IR  contribution:
\begin{equation}
Q_T(p)\propto
\int\limits_{-\infty}^{+\infty}\!\! dp_z 
\frac{p^2}{\alpha^2}
\frac{(p_z^2+p^2)^{3/2}}{(p_z^2\!+\!p^4/(4\alpha^2))^2}.
\label{LoopCorrKinkNaive}
\end{equation}
This integral is saturated in IR region
 at $p_z \sim p^2/\alpha $, hence
\begin{equation}
Q_T(p) \propto \frac{p^2}{\alpha^2}
 \frac{(p^2/\alpha) p^3}{(p^8/\alpha^4)}
\propto \frac{\alpha}{p},
\label{naivePiabMkink}
\end{equation}
which coincides with Eq.~(\ref{PiTransverse}). 
  Therefore, the one loop contribution of massless fermions 
  to the vector brane-to-brane propagator is
\begin{equation}
G_{\mu\nu}^{(1)}(p)= \frac{3 (\g_4)^2 }{ 2^{9} \pi } 
\frac{\alpha^2}{p^3}
\left(\eta_{\mu\nu}-\frac{p_{\mu}p_\nu}{p^2}\right),
\label{G1DWmslss}
\end{equation}
while the tree level propagator
 $G_A(p,0,0)$ is propotional to
 $1/p^2$ (see Eq.~(\ref{zeromodeVecKink})). This means 
 that the model  
 is inconsistent in the IR limit. One way to interpret the 
 result (\ref{G1DWmslss}) 
 is to claim that the theory is in the strong 
 coupling regime at $(\g_4)^2\alpha/p \lesssim 1$, 
 where the one loop correction exceeds the tree level
 propagator. We conclude that the model with mass gap between
 zero and massive vector KK modes is pathological in 
 IR, as expected.

\section{ RSII-1 set up
\label{RSII-1}}
In this section we consider a model
without mass gap between zero and massive vector KK 
modes.
Namely, we calculate the
 one loop contribution
of massles fermions to the brane-to-brane vector propagator 
in the framework of RSII-1 model with one compact and 
one infinite extra dimensions.  
6D  mectric of Euclidean RSII-1 
 model is
\begin{equation}
\begin{split}
ds^2&=G_{MN}dx^Mdx^N
\\
&=w^2(z)(\eta_{\mu\nu} dx^\mu dx^\nu+d\theta^2+dz^2),
\label{ads6Metric}
\end{split}
\end{equation}
where indices $M,N$ denote the coordinates of 
6D spacetime,~$M,N=0,1,2,3,5,6$. Greek indices 
   $\mu,\nu$ label 4D subspace, 
  $\mu,\nu=0,1,2,3$, $x^5$ is the compact extra dimension,  
$x^5\equiv 
\theta \in 
[0,2\pi R] $, and
$x^6\equiv z$  refers to extra dimension of infinite size
 $z\in[-\infty,+\infty]$.
The warp factor $w(z)$ is given by 
\begin{equation}
w(z)=1/(1+k|z|).
\label{warpRS}
\end{equation} 
From geometric point of view, the metric 
(\ref{ads6Metric}) describes 5-brane with one compact
dimension located at the point $z=0$
of bulk space. Let us consider the action 
of $U(1)$ gauge theory with massles fermions in the background 
(\ref{ads6Metric})
 \begin{equation}
 \begin{split}
 S_{RS}[\psi,A]=
 \int d^4 x\, dz\,
d\theta \sqrt{G}\Big[\frac{1}{4}F_{MN}F^{MN}+
\\
 + \,i\overline{\psi}\Gamma^{\bar{M}}E_{\bar{M}}^M(\nabla_M-i 
 \g_6 
 A_M)\psi \Big], 
 \label{U(1)action}
 \end{split}
 \end{equation}
where indices 
$\bar{M},\bar{N}$ label the tangent space and
$\nabla_M$ is spinorial covariant derivative.
Couplings and fields 
have the following mass dimensions:
 $[\g_6]=-1$, $[A_M]=2$ and
$[\Psi]=5/2$. The size of compact extra dimension $R$
is assumed to be  $R \ll 1/E $, where $E$ is the energy of 
interest.  In
  Secs.~\ref{VecGreenRSII} and~\ref{mslsRSII}  we study
KK excitations of fermions and bosons which are 
homogeneous along the compact extra dimension $\theta$. 
We discuss 
 in Sec.~\ref{KKRSII} 
  quantum corrections to the   photon 
propagator coming from the fermion states 
inhomogeneous  along~$\theta$.
 
\subsection{Vector field propagator in the RSII-1 set up
\label{VecGreenRSII}}
 \begin{figure}[t]
\begin{center}
\includegraphics[width=0.5\textwidth]{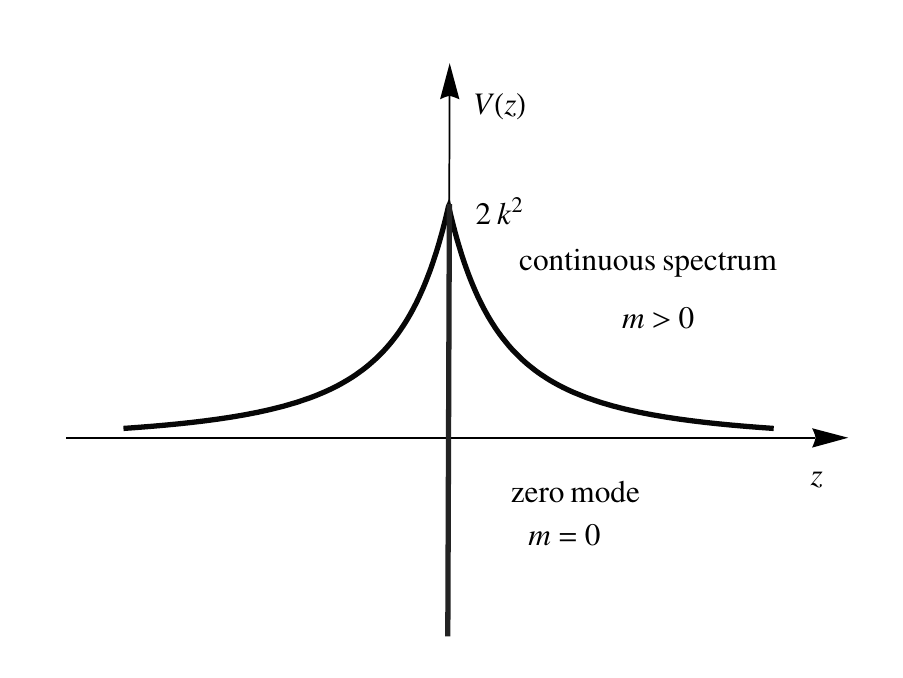}
\caption {RSII-1 effective potential.
 \label{figure4}}
\end{center}
\end{figure} 
 Let us find brane-to-bulk vector Green's function.
The vector part of the
 action  (\ref{U(1)action}) is
 \begin{equation}
 S[A]=\int d^4x\, dz\, d\theta \, w^2(z)\,\frac{1}{4} 
 F_{MN}^2, \label{VecAct}
 \end{equation}
where indices $M,N$ are contracted with flat metric. 
 The action~(\ref{VecAct})
  is analogous to that considered in domain wall set up
  (see vector part of the action~(\ref{DWaction})) 
  expect for the form of
 the warp factor $w(z)$.  
 We introduce the vector field $B_M$ in the same 
 way as in Sec.~\ref{theDWmod}
   (see Eq. (\ref{VecRescKink}))
\begin{equation}
B_M=w A_M.
\label{VecResc}
\end{equation}
We consider the field $B_M$ independent of $\theta$
with $B_\theta=0$ and choose the gauge $B_z=0$. 

 The equation of motion for KK mode $B^{(m)}(z)$ 
 of the field $B_\mu(x,z)$ coincides with 
 Eq.~(\ref{VecEq1}) up to  redefinition
  $\phi(z)\rightarrow w(z)$. In RSII-1 set
   up the spectrum is determined 
 by quantum - mechanical potential (see Fig.~\ref{figure4})
 \begin{equation}
 V(z)=\frac{w''}{w}=\frac{2k^2}{(1\!+\!k|z|)^2}-2k\,
\delta(z).
\label{VecPotRS}
 \end{equation}
Vector field $B^{(m)}(z)$ has a zero 
mode 
\begin{equation}
B^{(0)}(z)=\sqrt{\frac{k}{2}}\frac{1}{(1\!+\!k|z|)},
\end{equation} 
which has to do the delta function well in $V(z)$.
This zero mode corresponds to the constant 
field~$A^{(0)}(z)=\sqrt{k/2}$. The latter is homogeneous
along the large extra dimension $z$ and represents 
the photon localized on the brane.
In contrast to the domain wall set up, there is no 
mass gap, separating the zero mode $B^{(0)}(z)$ from  
the continuum of states~$B^{(m)}(z)$. 

  The brane-to-bulk vector Green's 
function 
$G^B_{\mu\nu}(p,z,0)=
\langle B_\mu(p,z)B_\nu(p,0)\rangle$
obeys 
\begin{equation}
\left(p^2\!-\!\partial_z^2\! +V(z)\right)G^B_{\mu\nu}(p,z,0)=\eta_{\mu\nu}\delta(z),\label{VecBraneToBulkGreenf}
\end{equation}
We obtain the solution to Eq.~(\ref{VecBraneToBulkGreenf}) 
in Appendix~\ref{RSIInProp}. It is given by 
$G^B_{\mu\nu}(p,z,0)=\eta_{\mu\nu}G_B(p,z,0)$, where 
\begin{equation}
G_B(p,z,0)=\frac{\e^{-p|z|}}{2p}+
\!\frac{k \e^{-p|z|}}{2p^2}\frac{1}{(1\!+\!k|z|)}.
\label{GreenVecPZ}
\end{equation}
The first term in 
Eq.~(\ref{GreenVecPZ}) is the Green's function 
of massles
 field in flat 5D spacetime. 
The second term of
Eq.~(\ref{GreenVecPZ}) is proportional to $1/p^2$, therefore
 this is the zero mode contribution to the  vector propagator
 at the tree
 level. In the IR regime $p/k \ll 1$, 
  brane-to-bulk propagator of the field $A_{\mu}$ has 
  the form
  \begin{equation}
G_A(p,z,0)\equiv G_B(p,z,0)/w(z)= \frac{k}{2p^2}\exp(-p|z|).  
\label{RSIInAZm}
  \end{equation}
  In contrast to the domain wall 
  case (see Eq.~(\ref{VecGFDWA})) $G_A(p,z,0)$
  of RSII-1 set up decreases more
  rapidly towards $z\rightarrow \infty$ in IR limit. 
  It is worth rewriting the
 Green's function  
 $G_A(p,z,0)$ in the momentum 
 space:
\begin{equation}
\widetilde{G}_A(p,p_z)\!=\!\!\int\limits_{-\infty}^{\infty}
\! dz\, 
 G_A(p,z,0) e^{i p_z z}\!=\frac{1}{P^2}+\frac{2p 
k}{P^4},  
\label{GreenFourierRS} 
\end{equation}
where $P$ is defined by Eq.~(\ref{5dPsquared}).

\subsection{One loop contribution of 
$\theta$-homogeneous fermions 
to the vector brane-to-brane propagator. \label{mslsRSII}}

In this section we derive one loop fermion correction 
to the vector Green's function from brane-to-brane.
It is known that massless fermions are conformal, i.~e.,
upon rescaling of the fermion field
\begin{equation}
\psi = w^{-5/2} \Psi,
\label{SpinorResc}
\end{equation} 
the fermion action reduces to the flat-space form
\begin{equation}
S[\Psi]\!=\!\!\int\!\! d^4 x\, dz\, d\theta \!\left(
i\overline{\Psi}\Gamma^{\mu}\partial_\mu \Psi
\!+\!i\overline{\Psi}\Gamma^{\theta}\partial_{\theta} \Psi
\!+\!i\overline{\Psi}\Gamma^{z}\partial_{z}\Psi
\right).
\label{FermRescAct}
\end{equation} 
where $\Gamma^\mu$, $\Gamma^\theta$ and $\Gamma^z$ are
Euclidean 6D gamma matrices; $\Psi$ is 
eight-component Dirac spinor 
$$\Psi=\left(\Psi_+,\Psi_-\right),$$
where $\Psi_+$ and $\Psi_-$ 
are four-component spinors which have appropriate
signs of 6D chirality. 
Consider now the KK mode expansion
\begin{equation}
 \Psi(x,z,\theta)= \frac{1}{\sqrt{2\pi R}}\Psi^{0}(x,z)+
 \sum_{n\ne 0} \Psi^{n}(x,z) f_S^{n}(\theta).
 \label{KKFermionsRS}
\end{equation}
where
 $\Psi^{0}(x,z)$ 
and $\Psi^{n}(x,z)f_S^{n}(\theta)$
 are $\theta$-homogeneous and $\theta$-inhomogeneous
KK modes, respectively. 
 We have dropped 6D
chirality indices $(\pm)$ in Eq.~(\ref{KKFermionsRS})
 for simplicity.
   Integrating out the compact 
extra dimension, one obtains
\begin{equation}
\begin{split}
 S[\Psi]\!=\!\int\!\! d^4x dz d \theta \sum_{n}
 \Big(i \overline{\Psi}^{n} \Gamma^\mu \partial_\mu \Psi^{n}+
\\   
   +\ 
 i\overline{\Psi} ^{n} \Gamma^z \partial_z
   \Psi^{n} + i m_n \overline{\Psi}^{n} 
   \Psi^{n}\Big),
   \end{split}
 \end{equation}
where $m_n=n/R$ is the mass of KK state.

Let us consider the one loop correction  to 
the vector propagator coming from the $\theta$-homogeneous
state $\Psi^{0}$. 
The interaction of $\Psi^{0}(x,z)$ and $A_{\mu}(x,z)$
is described by the action
\begin{equation}
S[A,\Psi^{0}]\!=\!\!\int\!\! d^4x\, dz \,
\g_5 \,
\overline{\Psi}^{0}\!(x,z) \Gamma^\mu 
A_\mu(x,z)\Psi^0(x,z),
\label{intzeroRS}
\end{equation}
where $\g_5$ is the 5D coupling
\begin{equation}
\g_5=\frac{\g_6}{\sqrt{2\pi\!R}}
\label{l5l6}
\end{equation}
 with mass dimension $[\g_5]=-1/2$.
 
The action (\ref{intzeroRS}) is analogous
 to the vector-fermion
 coupling in the domain wall set up (see 
 Eq.~\ref{IntDomainWall}). 
Since there are two types of four-component 
Dirac spinors in 6D model, one can carry over
 Eqs.~(\ref{IntDomainWall})~-~(\ref{QsDW}) 
to the RSII-1 scenario up to the 
fermion doubling factor in Eq.~(\ref{PiABFourier})
$$\widetilde{\Pi}_{\mu\nu}(p,p_z) \rightarrow 2 
\widetilde{\Pi}_{\mu\nu}(p,p_z).$$
Upon   substituting  Eqs.~(\ref{GreenFourierRS})
 and~(\ref{PiABFourier}) into Eq.~(\ref{LoopCorrection11}),
  one gets 
\begin{equation}
\begin{split}
G_{\mu\nu}^{(1)}(p)&=\!\frac{(\g_4)^2}{2\pi}
\frac{3\pi}{64}\,
\frac{2^3}{(4\pi)^2}
\frac{k}{2 p^2}
\Big[\left(\!\eta_{\mu\nu}\!-\!\frac{p_\mu p_\nu}{p^2}\!
\right) 
\!Q_T(p)+ 
\\
&+\frac{p_\mu p_\nu}{p^2}\, Q_L(p) \Big],
\label{PiABLoopCorrRS}
\end{split}
\end{equation}
where 5D and 4D effective couplings are related by 
Eq.~(\ref{g5g4}) up to the 
redefinition $\alpha \rightarrow k$.
 The functions $Q_T(p)$ and $Q_L(p)$ 
 are defined by Eq.~(\ref{QsDW}) 
 with $\widetilde{G}(p, p_z)$ given by
  Eq.~(\ref{GreenFourierRS})
 . Then, after integrating 
 Eq.~(\ref{QsDW}) over $p_z$, one finds
\begin{equation}
\begin{split}
&Q_T(p)=\!\frac{32}{3} +
\frac{16\, p}{k}+\frac{4p^2}{k^2}
\ln \frac{2 \Lambda}{p}, 
\\
&Q_L(p)\!=\!\frac{32}{15} +\frac{16}{3} 
\frac{p}{k}+\frac{4 p^2}{k^2}
\left(\!\ln \frac{2 \Lambda}{p}-1\right),
\label{QRSII1}
\end{split}
\end{equation}
Therefore, the one loop fermion 
contribution $G^{(1)}_{\mu\nu}(p)$
to the vector brane-to-brane 
propagator $G^{(0)}_{\mu\nu}(p)$ (cf. 
Eq.~(\ref{GreenVecPZ}) at 
$z=0$) in IR regime
 is given by
\begin{equation}
G^{(1)}_{\mu\nu}(p)=\frac{(\g_4)^2}{  (4\pi)^2} 
\frac{k}{p^2}\left(\eta_{\mu\nu}-
\frac{4}{5} \frac{p_{\mu}p_{\nu}}{p^2}\right).
\label{G1RS}
\end{equation}
It is worth noting that both $G^{(1)}_{\mu\nu}(p)$ and
$G^{(0)}_{\mu\nu}(p)$ are proportional to $k/p^2$ as
$p\rightarrow 0$. This means that there is no 
IR singularity in the one loop vector brane-to-brane 
propagator
of the RSII-1 set up. This is in contrast to
Sec.~\ref{DomainWallAction}.

One can understand the IR behaviour of $G^{(1)}_{\mu\nu}(p)$
in a way analogous to that
considered in Sec.~\ref{DomainWallCorrections}.
Namely,  let us consider Eq.~(\ref{QsDW})
in the case of RSII-1 model. From
 (\ref{GreenFourierRS})
it follows that if $P \rightarrow 0$ then 
$\widetilde{G}_A(p,p_z)\propto kp/P^4$.
Hence the main IR contributions to $Q_T(p)$ and $Q_L(p)$ 
come from the integrals
\begin{equation}
\begin{split}
Q_T(p)
\propto \!\!\!
\int\limits_{-\infty}^{+\infty}\!\! dp_z \frac{p^2}{k^2} 
\!\!\left[\frac{k p}{(p_z^2+p^2)^2}\right]^2\!\!
(p_z^2+p^2)^{3/2}, \,\,
\\
Q_L(p)
\propto \!\!\!
\int\limits_{-\infty}^{+\infty}\!\! dp_z \frac{p^2}{k^2} 
\!\!\left[\frac{k p}{(p_z^2+p^2)^2}\right]^2\!\!(p_z^2+p^2)^{1/2} p_z^2,
\label{LoopCorrRSNaive}
\end{split}
\end{equation}
 These integrals are saturated 
 at $p_z \sim p$, therefore one has
\begin{equation}
Q_T(p) \propto Q_L(q)\propto p\, \frac{p^2}{k^2} 
\frac{k^2 p^2}{p^8} p^3 \propto O(1)
\end{equation}
which coincides with Eq.~(\ref{QRSII1}) for $p/k \ll 1$. 

Thus, the analysis of the one loop vector propagators in
the two brane world models shows
 that the IR behaviour of vector 
brane-to-brane propagators is in one-to-one 
correspondence with the existence of the mass gap. 
While there is IR pathology in the model with the gap, 
 the gapless model is IR healthy.

\subsection{Contribution of $\theta$-inhomogeneous 
 KK fermions.
\label{KKRSII}}
In this section we derive one loop contribution
of $\theta$-inhomogeneous fermions
 to the vector brane-to-brane
propagator.
 Since the masses of 
corresponding KK
 excitations are large, $m_n=n/R \gg p $, 
 production of heavy inhomogeneous fermion modes 
is forbidden at the tree-level.
 Nevertheless, these modes may 
 contribute to the vector brane-to-brane propagator
 at the one loop level. 
 
  Let us recall that RSII-1
model is a 6D non-renormalizable spinor QED 
 with warped extra dimension. If the set up were 
  6D spinor QED on flat background, we would have to
 impose the following condition on gauge 
 coupling $\g_6$ and UV cutoff scale $\Lambda$
\begin{equation}
 \Lambda \ll 1/\g_6.
 \label{couplingcutoff}
\end{equation}
 In warped space-time, it is appropriate to use
 the position dependent cutoff formalizm 
 \cite{Randall:2001gb}. Namely, the cutoff 
 scale at given $z$ 
 is $\Lambda_{\tt eff}(z)=\Lambda/(1\!+\! k z)$.
In particular, one imposes position
dependent upper bound on KK masses
\begin{equation}
m_n<\Lambda/(1\!+\! k z).
\label{KKupperbound}
\end{equation}
Thus, the one-loop 
contribution of $\theta$-inhomogeneous KK states can 
be written as follows:
\begin{equation}
\begin{split}
G_{\mu\nu}^{KK}(p)&=\!
(\g_5)^2\!\!\!
\int\limits_{-\infty}^{+\infty}\!\! dz_1 dz_2\, 
G_A(p,\!z_1,\!0)G_A(p,\!z_2,\!0) \times
\\
&\times
\int\limits_{-\infty}^{+\infty}\!\frac{dp_z}{2\pi} \, 
\e^{ip_z(z_1-z_2)}\!\!\!\!
 \sum_{n\ne 0}^{N_{KK}(z)}\!\!\! 
  \widetilde{\Pi}_{\mu\nu}^{n} (p, p_z),
\label{PiABmassive}
\end{split}
\end{equation}
where $N_{KK}(z)$ is the effective number of KK states, and 
$z=(z_1\!+\!z_2)/2$;
 $\widetilde{\Pi}_{\mu\nu}^{n} (p, p_z)$ is the
 one loop fermion integral of massive KK excitations 
(compare with massless case Eq.~(\ref{PiabtildeKink})
and (\ref{PiABFourier}))
 \begin{equation}
\widetilde{\Pi}^{n}_{\mu\nu}(p,p_z)\!=\!\!\!
\int\limits_0^1\!\! dx \frac{2^3}{(4\pi)^2} \, 2 
\sqrt{\Delta_n}\!
\left(\eta_{\mu\nu}P^2\!-\!p_\mu p_\nu\!\right)x(1-x),
\label{PiabtildeKink2} 
\end{equation}
where $\Delta_n=m_n^2+x(1-x)P^2$. Since $m_n \gg p$, we set 
$\Delta_n=m_n^2$ in Eq.~(\ref{PiabtildeKink2}), and then 
evaluate
the integral.  In this way we obtain 
 in IR regime $P \ll k$ 
  \begin{equation}
  \widetilde{\Pi}^{n}_{\mu\nu}(p,p_z)=
   \frac{2^3}{(4\pi)^2}\, 2 \,\frac{m_n}{6}
   \left(\eta_{\mu\nu}P^2 -p_\mu p_\nu\right),
  \end{equation}
  Let us use the variables $z=(z_1\!+\!z_2)/2$ and $z'=z_1\!-\!z_2$ in  
 Eq.~(\ref{PiABmassive}).
Implementing the cutoff condition (\ref{KKupperbound}), 
we interchange the order of bulk space integration 
and KK summation  in 
Eq.~(\ref{PiABmassive}). Then,  after integrating 
over $p_z$, we have
\begin{equation}
\begin{split}
&G^{KK}_{\mu\nu}(p)
\!=\!(\g_5)^2 \!\!\sum_{n\ne 0}^{N_{KK}} \!\!
\frac{m_n}{6}
 \!\!\!
\int\limits_{-z_{IR}^{n}}^{+z_{IR}^{n}}\!\!\!\!
 dz\, dz'\delta(z') \frac{2^3}{(4\pi)^2} 2\, \times
\\ 
 &\times
 \left[\left(\eta_{\mu\nu}p^2\!-\!p_\mu p_\nu\right)\!-\!
 \eta_{\mu\nu}\frac{\partial^2}{\partial z'^2}\right]\times
 \\
&\times G_A(p,z\!+\!z'/2,0) 
G_A(p,z\!-\!z'/2,0),
 \label{PiABmassive22}
 \end{split}
\end{equation}
where $z_{IR}^{n}=(\Lambda-m_{n})/(k m_{n})$, the masses
of KK states are bounded now by $m_n < \Lambda$. Hence, 
the number of KK states in (\ref{PiABmassive22}) is
\begin{equation}
N_{KK}=R \Lambda.
\end{equation}
Integrating Eq.~(\ref{PiABmassive22}) over $z'$ and $z$, 
one obtains
\begin{equation}
\begin{split}
G^{KK}_{\mu\nu}(p)\!= &\!
(\g_5)^2\!\!\sum_{n\ne0}^{N_{KK}}\!\!
 \frac{m_{n}}{6} \frac{2^3}{(4\pi)^2}
\!\Big(\left[\eta_{\mu\nu}\!-\!
\frac{p_\mu p_\nu}
{p^2}\right]\!\mathcal{F}^{n}_1(p)+
\\
&+\eta_{\mu\nu}
\mathcal{F}^{n}_2(p)\Big),
\end{split}
\end{equation}
where $\mathcal{F}^{n}_1(p)$ and $\mathcal{F}^{n}_2(p)$
are given by 
\begin{equation}
\begin{split}
\mathcal{F}^{n}_1(p)&=\frac{5}{8}\frac{k^2}{p^3}
(1-\e^{-p z_{IR}^{n}})+
\frac{3}{8}\frac{k}{p^2}\Big(2(1-\e^{-p z_{IR}^{n}})-
\\
& -\e^{-p z_{IR}^{n}} k z_{IR}^{n} 
\Big)+
\frac{1}{16 p}\Big( 4(1-\e^{-p z_{IR}^{n}})-
\\
&-
4\e^{-p z^n_{IR}} k z^n_{IR}-\e^{-p z_{IR}^{n}} 
(k z_{IR}^{n})^2\Big)
\label{FKK1p}
\end{split}
\end{equation}
\begin{equation}
\begin{split}
\mathcal{F}^{n}_2(p)&=\frac{k}{2p^2}\e^{-p z_{IR}^{n}}
\bigl(1+k z_{IR}^{n}/2 \bigr)+
\\
&+\frac{1}{2p}\e^{-p z_{IR}^{n}}
\bigl(1+k z_{IR}^{n}+(k z_{IR}^{n})^2/4\bigr).
\end{split}
\label{FKK2p}
\end{equation}
Let us consider low momentum regime 
 $p z_{IR}^{n}\ll 1$, then at $ p/k \ll 1$ 
one has
$$
\mathcal{F}_1^{n}(p)=\mathcal{F}_2^{n}(p)= 
\frac{z_{IR}^{n}}{4} \frac{k^2}{p^2}.
$$
Therefore
$$
G^{KK}_{\mu\nu}(p)= 
 \frac{(\g_4)^2}{48 \pi^2}\sum_{n\ne 0}^{N_{KK}}
  \frac{(\Lambda-m_{n})}{p^2}
\left(2\eta_{\mu\nu}-\frac{p_\mu p_\nu}{p^2}\right),
$$
which leads to 
\begin{equation}
G^{KK}_{\mu\nu}(p)= 
 \frac{(\g_4)^2}{48 \pi^2}
  \frac{\Lambda (\Lambda R+1)}{2p^2}
\left(2\eta_{\mu\nu}-\frac{p_\mu p_\nu}{p^2}\right)
\label{thetainhomcontrb}
\end{equation}
  It follows 
  from Eq.~(\ref{l5l6}) and Eq.~(\ref{g5g4}) that 
  6D and 4D couplings are related by
  $$
  \g_6=\g_4 \sqrt{\frac{2\pi R}{k}},
  $$
therefore  Eq.~(\ref{thetainhomcontrb})
 is finally written as
\begin{equation}
G^{KK}_{\mu\nu}(p)= 
 \frac{(\g_6 \Lambda)^2}{192 \pi^3}
  \frac{k}{p^2}
\left(2\eta_{\mu\nu}-\frac{p_\mu p_\nu}{p^2}\right).
\label{2thetainhomcontrb}
\end{equation}
This is indeed a small correction to the tree-level
propagator, provided that $(\g_6 \Lambda)^2 \ll 1$, 
which coincides with 
Eq.~(\ref{couplingcutoff}). We conclude that the
entire 
RSII-1 scenario is viable at least at the one loop level.

\section{Discussion and conclusion
}\label{discussion}
In this paper we have considered two brane-world 
models with different mechanisms of gauge field localization
on the brane. We found that the fermion one loop correction
to the vector brane-to-brane propagator has a 
pathological IR divergence in the framework of 
5D massles spinor QED with gauge field localized
 on the domain wall, which makes this 
 model inconsistent. This result is consistent 
  with the observation in 
 Ref.~\cite{Smolyakov:2011hv}. 
 We have also considered 6D massles spinor QED
 in the background of modified Randall-Sundrum metric. 
 We have explicitly calculated the
 one loop fermion cotribution  to the vector brane-to-brane 
 propagator in this framework in the low energy limit. This 
 contribution is healthy in IR, 
so one
 can  consider the RSII-1 set up as consistent brane world 
 scenario, at least at the one loop level.
 
 We conclude that IR are inherent in models 
 with gauge field zero mode separated from 
 heavier modes by a gap, while models without the gap
 may be healthy in IR.

\section{Acknowledgements}
We are indebted to
D.~S.~Gorbunov, A.~L.~Kataev,  M.~Y.~Kuznetsov, D.~G.~Levkov,   
A.~G.~Panin, V.~A.~Rubakov and M.~A.~Smolyakov 
for helpful  discussions.  
This  work  was supported in part by 
grants of  Russian Ministry of  Education 
and  Science NS-5590.2012.2 and
GK-8412, grants of the President of 
Russian Federation MK-2757.2012.2, and grants
of RFBR 12-02-31595 MOL A and
RFBR 13-02-01127 A. 

\appendix

\section{ Vector field propagator 
in the domain wall set up.}
\label{GreeDW}

In this appendix we derive the vector  propagator 
from the brane to bulk in the model of
Sec.~\ref{DomainWallAction}.
We take the solution to Eq.~(\ref{kinkVecGrF2}) 
in the  form 
$G^B_{\mu\nu}(p,z,0)=\eta_{\mu\nu}G_B(p,z,0)$, where
\begin{equation}
G_B(p,z,0)=\theta(z)G^{(+)}(p,z)+
\theta(-z)G^{(-)}(p,z).
\label{KinkVFGFanzats}
\end{equation}
Here  $G^{(+)}(p,z)$ and $G^{(-)}(p,z)$
are linear combinations of odd and even solutions
\begin{equation}
\begin{split}
G^{(+)}(p,z)=B^{(+)} G^{(\e)}(p,z)+C^{(+)} G^{(o)}(p,z),
\\
G^{(-)}(p,z)=B^{(-)} G^{(\e)}(p,z)+C^{(-)} G^{(o)}(p,z).
\end{split}
\label{KinkVecEvenOddExpansion}
\end{equation}
Here
\begin{equation}
G^{(\e)}(p,z)=\ch^2\alpha z\,\, _2F_1(a,b;
\frac{1}{2};-\sh^2 \alpha z)
\label{Geven}
\end{equation}
\begin{equation}
G^{(o)}(p,z)=\sh \alpha z \, \ch^2\alpha z\,\, 
_2F_1(a+\frac{1}{2},b+\frac{1}{2};\frac{3}{2};-\sh^2 \alpha z),
\label{Godd}
\end{equation}
where $_2F_1$ is hypergeometric function, parameters 
  $a$ and $b$ are defined by
\begin{equation}
a=1-\chi/(2 \alpha),
\quad
b=1+\chi/(2 \alpha),
\end{equation}
with
\begin{equation}
\chi=\sqrt{p^2+\alpha^2}.
\end{equation}
 We impose the boundary condition far from the brane 
\begin{equation}
G^{(+)}(p,z)\Big|_{z=\infty}=0, 
\quad G^{(-)}(p,z)\Big|_{z=-\infty}=0. 
\label{KinkVecBoundCond}
\end{equation}
It follows from 
Eqs.~(\ref{KinkVecEvenOddExpansion}) 
and~(\ref{KinkVecBoundCond}) that
\begin{equation}
B^{(+)}=-C^{(+)}\lim_{z\rightarrow \infty} 
\frac{G^{(o)}(p,z)}{G^{(\e)}(p,z)}\equiv-C^{(+)} D(p)
\label{D(p)def}
\end{equation}
\begin{equation}
\begin{split}
B^{(-)}&=-C^{(-)}\lim_{z\rightarrow -\infty} 
\frac{G^{(o)}(p,z)}{G^{(\e)}(p,z)}=
\\
&=C^{(-)}\lim_{z\rightarrow \infty} 
\frac{G^{(o)}(p,z)}{G^{(\e)}(p,z)}\equiv C^{(-)}D(p).
\end{split}
\end{equation}
Then Eq.~(\ref{KinkVecEvenOddExpansion}) can be written 
in the following form:
\begin{equation}
\begin{split}
G^{(+)}(p,z)=C^{(+)}[-D(p) G^{(\e)}(p,z)+ G^{(o)}(p,z)],
\\
G^{(-)}(p,z)=C^{(-)}[D(p) G^{(\e)}(p,z)+ G^{(o)}(p,z)].
\label{KinkVecEvenOddExpansion2}
\end{split}
\end{equation}
Matching condition $G^{(+)}(p,0)=G^{(-)}(p,0)$ at the brane position
 $z=0$ gives 
 $$C^{(+)}=-C^{(-)}\equiv C.$$
Discontinuity of the derivative
$$-\partial_z G^{(+)}(p,0)+\partial_z G^{(-)}(p,0)=1$$
 at the point $z=0$ yields 
 $$C=-1/(2 \alpha).$$ 
 Thus, the propagator
 $G_B(p,z,0)$ reads
\begin{equation}
G_B(p,z,0)=\frac{1}{2\alpha}[D(p)
G^{(\e)}(p,z)-G^{(o)}(p,|z|)].
\label{Gbrnblk}
\end{equation} 
Expanding 
$G^{(\e)}(p,z)$ and $G^{(o)}(p,z)$ at 
large values of the variable
$\xi=\ch \alpha z$, one has
\begin{equation}
\begin{split}
&G^{(o)}(p,\xi)\mbox{sign}(z)
\!=\!\frac{\xi^{\frac{\chi}{\alpha}}}{2}\!\!
\left(\! \frac{2^{\frac{\chi}{\alpha}} \alpha }{\alpha\!+\!\chi}
+ O\left(1/\xi^2\right)\right)
\\
&+\xi^{-\frac{\chi}{\alpha}}
\left( \frac{2^{-\frac{\chi}{\alpha}} \, \alpha }{\alpha-\chi}
+ O\left(1/\xi^2\right)\right),
\end{split}
\label{GoddLargeExpansion}
\end{equation}
\begin{equation}
\begin{split}
 &G^{(\e)}(p,\xi)
\!=\!\frac{\xi^{\frac{\chi}{\alpha}}}{2}
\!\!
\left(\frac{2^{\frac{\chi}{\alpha}}
(-\alpha\!+\!\chi)}{\chi}
+ O\left(1/\xi^2\right)\right)
\\
&+\xi^{-\frac{\chi}{\alpha}}
\Bigl( \frac{2^{-\frac{\chi}{\alpha}} \,(\alpha+\chi) }{\chi}
+ O(1/\xi^2)\Bigr).
\end{split}
\label{GevenLargeExpansion}
\end{equation}
This gives
\begin{equation}
D(p)=\lim_{\xi\rightarrow \infty}\frac{G^{(o)}(p,\xi)}{G^{(\e)}
(p,\xi)}=\frac{\alpha \chi}{(\chi^2-\alpha^2)}.
\label{DpDefinition}
\end{equation}
Substituting Eqs.~(\ref{Geven}), (\ref{Godd}) 
 and (\ref{DpDefinition})
into  Eq.~(\ref{Gbrnblk}), and using the identity 
\begin{equation}
\begin{split}
&\frac{1}{(y\!-\!1/y)}\,_2F_1\! \left(\!1\!-\!\frac{y}{2},\!1\!+\! 
\frac{y}{2};\!\frac{1}{2};
\!-\sh^2 t \right)\!
\\
& -
\sh |t| _2F_1\left(\frac{3}{2}\!-\!\frac{y}{2},
\frac{3}{2}+\frac{y}{2};\frac{3}{2 };-\sh^2 t\right)=
\\
&=\frac{1}{2\ch^3 t}\left(\frac{\e^{-(y-1)|t|}}{y\!-\!1}\!+\!
\frac{\e^{-(y+1)|t|}}{y\!+\!1}\right),
\end{split}
\end{equation}
we obtain the final form 
 of the brane-to-bulk vector propagator
\begin{equation}
 G_B(p,z,0)=
 \frac{1}{4\,\ch \alpha z}\left(\frac{\e^{-(\chi-\alpha)|z|}}
 {\chi-\alpha}+
 \frac{\e^{-(\chi+\alpha)|z|}}{\chi+\alpha}\right).
\end{equation}

\section{
 Vector field propagator 
in the RSII-1 set up \label{RSIInProp}}
In this appendix we derive brane-to-bulk vector 
propagator $G_B(p,z,0)$ in 
the framework of RSII-1. The latter obeys 
\begin{equation}
\left(p^2-\partial_z^2 +\frac{w''}{w}\right)
G_B(p,z,0)=\delta(z).
\label{RSGrEq1}
\end{equation}
We take the solution to Eq.~(\ref{RSGrEq1})
 in the following form 
$$
G_B(p,z,0)=\theta(z)G^{(+)}(p,z)+\theta(-z)G^{(-)}(p,z),
$$
where  
$$
G^{(\pm)}(p,z)\!=\!C^{(\pm)}\!\!\left(\!1\!+\!\frac{k}{p}
\frac{1}{(1\!\pm\!kz)}\right)
\!\e^{-\frac{p}{k}(1\!\pm\!kz)}.
$$
Matching condition of $G^{(+)}(p,z)$ and $G^{(-)}(p,z)$ 
and discontinuity of
derivative at the point $z=0$ are
$$
G^{(+)}(p,0)=G^{(-)}(p,0),
$$
$$
-\partial_z G^{(+)}(p,0)\!+\!
\partial_z G^{(-)}(p,0)\!-\!2k G(p,0)\!=\!1.
$$
This yields  
$$
C^{(+)}=C^{(-)}=\frac{1}{2p} \exp\left(\frac{p}{k}\right).
$$
Hence, we obtain
\begin{equation}
G_B(p,z,0)=\frac{\e^{-p|z|}}{2p}+
\!\frac{k \e^{-p|z|}}{2p^2}\frac{1}{(1\!+\!k|z|)}.
\end{equation}


\end{document}